\documentclass[11pt]{article}
\usepackage{amsmath}
\usepackage{amssymb}
\usepackage{amsthm} 
\usepackage{bbm}
\usepackage{graphicx}
\usepackage{algorithm}
\usepackage{algorithmic}
\usepackage{clrscode}
\usepackage[top=1in, bottom=1in, left=1in, right=1in]{geometry}

\newtheorem{thm}{Theorem}
\newtheorem{lem}{Lemma}

\newtheorem{rem}{\it Remark}

\DeclareMathOperator*{\argmin}{arg\,min}

\begin{document}
\pagestyle{plain}
\newenvironment{frcseries}{\fontfamily{frc} \selectfont}{}
\newcommand{\textfrc}[1]{{\frcseries #1}}
\newcommand{\mathfrc}[1]{\text{\textfrc{#1}}}

\title{Fast Subspace Approximation via Greedy Least-Squares}
%
\author{M. A. Iwen\thanks{Contact Author.  Supported in part by NSA grant H98230-13-1-0275.} \\
Department of Mathematics, Michigan State University \\
Department of Electrical and Computer Engineering, Michigan State University \\
\textit{Email:}  markiwen@math.msu.edu\\
\and\\ Felix Krahmer\\
Institute for Numerical and Applied Mathematics, University of G\"ottingen\\
\textit{Email:}  f.krahmer@math.uni-goettingen.de\\
}

\maketitle

\begin{abstract}
In this note, we develop fast and deterministic dimensionality reduction techniques for a family of subspace approximation problems.  Let $P \subset \mathbbm{R}^N$ be a given set of $M$ points.  The techniques developed herein find an $O(n \log M)$-dimensional subspace that is guaranteed to always contain a near-best fit $n$-dimensional hyperplane $\mathcal{H}$ for $P$ with respect to the cumulative projection error $\left( \sum_{{\bf x} \in P} \| {\bf x} - \Pi_\mathcal{H} {\bf x} \|^p_2 \right)^{1/p}$, for any chosen $p > 2$.  The deterministic algorithm runs in $\tilde{O} \left( MN^2 \right)$-time, and can be randomized to run in only $\tilde{O} \left( MNn \right)$-time while maintaining its error guarantees with high probability.  In the case $p = \infty$ the dimensionality reduction techniques can be combined with efficient algorithms for computing the John ellipsoid of a data set in order to produce an $n$-dimensional subspace whose maximum $\ell_2$-distance to any point in the convex hull of $P$ is 
minimized.  The resulting algorithm remains $\tilde{O} \left( MNn \right)$-time.  In addition, the dimensionality reduction techniques developed herein can also be combined with other existing subspace approximation algorithms for $2 < p \leq \infty$ -- including more accurate algorithms based on convex programming relaxations -- in order to reduce their runtimes.
\end{abstract}


\section{Introduction}
\label{sec:intro}

Fitting a given point cloud with a low-dimensional affine subspace is a fundamental computational task in data analysis.  In this paper we consider fast algorithms for approximating a given set of $M$ points, $P \subset \mathbbm{R}^N$, with an $n$-dimensional affine subspace $\mathcal{A} \subset \mathbbm{R}^N$ that is a near-best fit.   Here the fitness of $\mathcal{A}$ will be measured by $d^{(p)}(P,\mathcal{A}) := \sqrt[p]{\sum_{{\bf x} \in P} \left( d({\bf x},\mathcal{A}) \right)^p}$,  
where $d({\bf x},\mathcal{A})$ is the Euclidean distance from ${\bf x}$ to $\mathcal{A}$, and $p \in \mathbbm{R}^+$.  Similarly, when $p = \infty$ the fitness measure will be $d^{(\infty)}(P,\mathcal{A}) := \max_{{\bf x} \in P} d({\bf x},\mathcal{A})$.  An $n$-dimensional affine subspace $\mathcal{A} \subset \mathbbm{R}^N$ is a \textit{near-best fit for $P$} with respect to this fitness measure if there exists a small constant $C \in \mathbbm{R}^+$ such that $d^{(p)}(P,\mathcal{A}) \leq C \cdot d^{(p)}(P,\mathcal{H})$ for all $n$-dimensional affine subspaces $\mathcal{H} \subset \mathbbm{R}^N$.\footnote{The approximation constant $C$ may depend (mildly) on both $p$ and $|P| = M$.}  In this paper we are interested in calculating near-best fit affine subspaces for large and high-dimensional point sets, $P \subset \mathbbm{R}^N$, as rapidly as possible.

In the case $p = 2$ the problem above is the well known least-squares approximation problem.  Mathematically, a near-best fit $n$-dimensional least-squares subspace can be obtained by computing the top $n$ eigenvectors of $XX^{\rm T}$ for the matrix $X \in \mathbbm{R}^{N \times M}$ whose columns are the points in $P$.  Decades of progress related to the computational eigenvector problem has resulted in many efficient numerical schemes for this problem (see, e.g., \cite{watkins2007matrix,halko2011finding}, and the references therein).  The situation is more difficult when $p \neq 2$.  None the less, a good deal of work has been done developing algorithms for other values of $p$ as well.

Examples include methods for approximately solving the case $p = 1$, which has been proposed as a means of reducing the effects of statistical outliers on an approximating subspace (see, e.g., \cite{DBLP:journals/corr/abs-1202-4044}).  However, in this paper we are primarily interested in $p > 2$.  In particular, we develop fast dimensionality reduction techniques for the subspace approximation problem which can be used in combination with existing solution methods for any $p > 2$ \cite{shyamalkumar2007efficient,deshpande2011algorithms} in order to reduce their runtimes.  For the important case $p = \infty$ these new dimensionality reduction methods yield a new fast approximation algorithm guaranteed to find near-optimal solutions.  

\subsection{Results and Previous Work for the $p = \infty$ Case}

The case $p = \infty$ is closely related to several fundamental computational problems in convex geometry and has been widely studied (see, e.g., \cite{gritzmann1994complexity,faigle1996note,har2002projective, ye2003improved,agarwal2005geometric,varadarajan2007approximating}, and references therein).  Previous computational methods developed for this case can be grouped into two general categories:  methods based on semi-definite programming relaxations (e.g., \cite{ye2003improved,varadarajan2007approximating}), and methods based on core-set techniques (e.g., \cite{har2002projective,agarwal2005geometric}).  Both approaches have comparative strengths.  The semidefinite programming approach leads to highly accurate approximations.  In particular, \cite{varadarajan2007approximating} demonstrates a randomized approach which computes an $n$-dimensional subspace $\mathcal{A}$ that has $d^{(\infty)}(P,\mathcal{A}) \leq \sqrt{12 \log M} \cdot d^{(\infty)}(P,\mathcal{H})$ for all $n$-dimensional subspaces $\mathcal{H}
 \subset \mathbbm{R}^N$ with high probability.  Furthermore, the approximation factor $\sqrt{12 \log M}$ is shown to be close to the best achievable in polynomial time.  However, the method requires the solution of a semi-definite program, and so has a runtime complexity that scales super-linearly in both $M$ and $N$.  This makes the technique intractable for large sets of points in high dimensional space.  

The core-set approach achieves better runtime complexities for small values of $n$.  In \cite{agarwal2005geometric} a $\tilde{O}(M N 2^{n})$-time randomized approximation algorithm is developed for the $p = \infty$ case.\footnote{Herein, $\tilde{O}(\cdot)$-notation indicates that polylogarithmic factors have been dropped from the associated $O$-upper bounds for the sake of readability.}  This algorithm has the advantage of being linear in both $M$ and $N$, but quickly becomes computationally infeasible as the dimension of the approximating subspace, $n$, grows.  

In this paper we develop an $\tilde{O}(M N^2)$-time deterministic algorithm which computes an $n$-dimensional subspace $\mathcal{A}$ that is guaranteed to have $d^{(\infty)}(P,\mathcal{A}) \leq C \sqrt{n \log M} \cdot d^{(\infty)}(P,\mathcal{H})$ for all $n$-dimensional subspaces $\mathcal{H} \subset \mathbbm{R}^N$.  Here $C \in \mathbbm{R}^+$ is a small universal constant (e.g., it can be made less than $10$).  Furthermore, the algorithm can be randomized to run in only $\tilde{O}(MNn)$-time while still achieving the same accuracy guarantee with high probability.  This improves on the runtime complexities of existing core-set approaches while simultaneously obtaining accuracies on the order of existing semi-definite programming methods for small $n$.

The approximation algorithms for the $p = \infty$ case developed in this paper are motivated by the following idea:  The difficulty of approximating $P \subset \mathbbm{R}^N$ with a subspace can be greatly reduced by first approximating (the convex hull of) $P$ with an ellipsoid, and then approximating the resulting ellipsoid with an $n$-dimensional subspace.  In fact, fast algorithms for approximating (the convex hull of) $P$ by an ellipsoid are already known (see, e.g., \cite{khachiyan1996rounding,kumar2005minimum,Todd20071731}).  And, it is straightforward to approximate an ellipsoid optimally with an $n$-dimensional subspace -- one may simply use its $n$ largest semi-axes as a basis.  The only deficit in this simple approach is that the accuracy it guarantees is rather poor.  The resulting $n$-dimensional subspace $\mathcal{A}$ may have $d^{(\infty)}(P,\mathcal{A})$ as large as $\sqrt{N} \cdot d^{(\infty)}(P,\mathcal{H})$ for some other $n$-dimensional subspace $\mathcal{H} \subset \mathbbm{R}^N$.  This 
guarantee can be improved, however, if $N$ (i.e., the dimension of the point set $P$) is reduced before the approximating ellipsoid is computed.  Motivated by this idea, we develop new dimensionality reduction algorithms for the subspace approximation problem below.

\subsection{Dimensionality Reduction Results and Previous Work}

An algorithm is a dimensionality reduction method for the subspace approximation problem if, for any $P \subset \mathbbm{R}^N$, it finds a low-dimensional subspace that is guaranteed to contain a near-best fit $n$-dimensional hyperplane $\mathcal{H}$.  Such dimensionality reduction methods can be regarded as a ``weak'' approximate solution methods for the subspace approximation problem in the following sense. They produce subspaces whose dimensions are larger than $n$ (i.e., larger than the target dimension of the desired best-fit hyperplane), but solving the problem restricted to these subspaces will yield a near-optimal solution.  Thus dimensionality reduction methods -- when sufficiently fast -- allow the subspace approximation problem to be simplified before more time intensive solution methods are employed.  For example, if a low-dimensional subspace has been found, which still contains a near-best fit solution, high-dimensional data (i.e., with $N$ large) can be projected onto that subspace in order to 
reduce its complexity before solving.  Hence, fast dimensionality reduction algorithms can be used to help speed up existing solutions methods for $p > 2$ (e.g., by reducing the input problem sizes for methods based on solving convex programs \cite{deshpande2011algorithms}.)

Several dimensionality reduction techniques have been developed for the subspace approximation problem over the past several years (see, e.g., \cite{agarwal2005geometric,deshpande2007sampling,feldman2011unified} and references therein).  These methods are all based on sampling techniques and either have runtime complexities that scale exponentially in $n$, or embedding subspace dimensions that scale exponentially in $p$.  In \cite{deshpande2007sampling}, for example, an $MN n^{O(1)}$-time randomized algorithm is given which is guaranteed, with high probability, to return an $\tilde{O}(n^{p + 3})$-dimensional subspace that itself contains another $n$-dimensional subspace, $\mathcal{A}$, whose fit, $d^{(p)}(P,\mathcal{A})$, is the near-best possible for any $p \in [1,\infty)$.  Although useful for small $p$, these methods quickly become infeasible as $p$ increases.

In this paper a different dimensionality reduction approach is taken that reduces the problem, for any $p \geq 2$, to a small number of least-squares problems.  The idea is to greedily approximate a large portion of the input data $P$ with a fast least-squares method.  It turns out that a large portion of $P$ is always well-approximated, \textit{for any $p > 2$}, by $P$'s best-fit $n$-dimensional least-squares subspace.  Then, the previously worst-approximated points in $P$ can be iteratively fit by least-squares subspaces until all of $P$ has eventually been approximated well, with respect to any desired $p > 2$, by the union of $O(\log M)$ least-squares subspaces.  Using this idea, a deterministic $\tilde{O}(M N^2)$-time algorithm can be developed which is always guaranteed to return an  $O(n \log M)$-dimensional subspace that itself contains another $n$-dimensional subspace, $\mathcal{A}$, whose fit, $d^{(p)}(P,\mathcal{A})$, is the near-best possible for any $p \in [2,\infty]$.  Furthermore, this 
algorithm can be randomized to run in only $\tilde{O}(MNn)$-time while still achieving the same accuracy guarantees as the deterministic variant with high probability.  

\subsection{Organization}
The remainder of this paper is organized as follows:  In Section~\ref{sec:Prelim} notation is established and necessary theory is reviewed.  Then, in Section~\ref{sec:dimReduct}, the dimensionality reduction results are developed for any $p>2$.  Finally, in Section~\ref{sec:CaseInfRes}, our improved dimensionality reduction result for the case $p = \infty$ is used to illustrate a fast and simple subspace approximation algorithm for the $p = \infty$ subspace approximation problem.

\section{Preliminaries:  Notation and Setup}
\label{sec:Prelim}

For any matrix $X \in \mathbbm{R}^{N \times M}$ we will denote the $j^{\rm th}$ column of $X$ by ${\bf X}_j \in \mathbbm{R}^{N}$.  The transpose of a matrix, $X \in \mathbbm{R}^{N \times M}$, will be denoted by $X^{\rm T} \in \mathbbm{R}^{M \times N}$, and the singular values of any matrix $X \in \mathbbm{R}^{N \times M}$ will always be ordered  as $\sigma_1(X) \geq \sigma_2(X) \geq \dots \geq \sigma_{\min(N,M)}(X) \geq 0.$  The Frobenius norm of $X \in \mathbbm{R}^{N \times M}$ is defined as
\begin{equation}
\| X \|_F := \sqrt{\sum^M_{j = 1} \sum^N_{i = 1} |X_{i,j}|^2} = \sqrt{\sum^{\min(N,M)}_{l=1} \sigma^2_l(X)}.
\end{equation}
A key ingredient of our results is the following perturbation bounds for singular values (see, e.g., \cite{HornJohnson1991Topics}).
\begin{thm}[Weyl]
Let $A, B \in \mathbbm{R}^{M \times N}$, and $q = \min \{ M, N \}$.  Then, 
$$\sigma_{i + j -1} (A + B) \leq \sigma_i (A) + \sigma_j (B)$$
holds for all $i,j \in \{ 1, \dots, q\}$ with $i + j \leq q+1$.
\label{Thm:WBounds}
\end{thm}

Given an $\tilde{n}$-dimensional subspace $\mathcal{S} \subseteq \mathbbm{R}^N$, we will denote the set of all $n$-dimensional affine subspaces of $\mathcal{S}$ by $\Gamma_{n} \left( \mathcal{S} \right)$.  Here, of course, we assume that $N \geq \tilde{n} \geq n$.  Given an affine subspace $\mathcal{A} \in \Gamma_{n} \left( \mathcal{S} \right)$, we will denote the offset of $\mathcal{A}$ by 
\begin{equation}
{\bf a}_{\mathcal A} := \argmin_{{\bf x} \in \mathcal{A}} \| {\bf x} \|_2,
\end{equation}
and the $n$-dimensional subspace of $\mathcal{S}$ that is parallel to $\mathcal{A}$ by
\begin{equation}
\mathcal{S}_{\mathcal A} := {\mathcal A} - {\bf a}_{\mathcal A} := \left\{ {\bf x} - {\bf a}_{\mathcal A}  ~\big |~ {\bf x} \in \mathcal{A} \right\}.
\end{equation}
Note that ${\bf a}_{\mathcal A} \in \mathcal{S}_{\mathcal A}^\perp$.  Thus, we may define the projection operator onto $\mathcal{A}$, $\Pi_{\mathcal A} : \mathbbm{R}^N \rightarrow \mathcal{A}$, by 
\begin{equation}
\Pi_{\mathcal A} {\bf x} := \Pi_{\mathcal{S_A}} {\bf x} + {\bf a}_{\mathcal A}.
\end{equation}
Here $\Pi_{\mathcal{S_A}}$ is the orthogonal projection onto $\mathcal{S_A}$.

\subsection{A Family of Distances}

Given a subset $T \subset \mathbbm{R}^N$ and an affine subspace $\mathcal{A} \in \Gamma_n (\mathcal{S})$ we will want to consider the ``distance'' of $T$ from $\mathcal{A}$, defined by
\begin{equation}
d^{(\infty)}(T,\mathcal{A}) := \sup_{{\bf x} \in T} \| {\bf x} - \Pi_{\mathcal A} {\bf x} \|_2.
\end{equation}
Let $\mathcal{S}$ be an $\tilde{n} \geq n$ subspace of $\mathbbm{R}^N$.  We can now define the Euclidean Kolmogorov $n$-width of $T$ in this setting by
\begin{equation}
d^{(\infty)}_n (T,{\mathcal S}) := \inf_{\mathcal{A} \in \Gamma_n (\mathcal{S})} ~ d^{(\infty)}(T,\mathcal{A}) = \inf_{\mathcal{A} \in \Gamma_n (\mathcal{S})} ~ \sup_{{\bf x} \in T} ~ \| {\bf x} - \Pi_{\mathcal A} {\bf x} \|_2.
\label{def:EKnW}
\end{equation}
Finally, we note that there will always be (at least one) optimal affine subspace, $\mathcal{A}_{\rm opt} \in  \Gamma_n (\mathcal{S})$, with
\begin{equation}
d^{(\infty)}(T,\mathcal{A}_{\rm opt}) = d^{(\infty)}_n (T,{\mathcal S})
\end{equation}
when $T$ is ``sufficiently nice'' (e.g., when $T$ is either finite, or convex and compact).\footnote{This follows from the fact that Stiefel manifolds are compact, together with the fact that only offsets, ${\bf a}_{\mathcal A} \in \mathbbm{R}^N$, contained in the ball of radius $\sup_{{\bf x} \in T} \| x \|_2$ are ever relevant to minimizing $d^{(\infty)}(T,\cdot)$.  Thus, the set of relevant affine subspaces under consideration is compact when $T$ is bounded.  Finally, $d^{(\infty)}(T,\cdot ):  \Gamma_{n} \left( \mathcal{S} \right) \rightarrow \mathbbm{R}^+$, $T \subset \mathbbm{R}^N$ fixed, will be continuous when $T$ is sufficiently well behaved (e.g., either finite, or compact and convex).}

When $T = \{ {\bf t}_1, \dots, {\bf t}_M \} \subset \mathbbm{R}^N$ is finite, we may define a vector ${\bf e}_{\mathcal{A}} \in \mathbbm{R}^M$ for any given $\mathcal{A} \in \Gamma_n \left( \mathbbm{R}^N \right)$ by
\begin{equation}
\left( {\bf e}_{\mathcal{A}} \right)_j := \left \| {\bf t}_j - \Pi_{\mathcal A} {\bf t}_j \right \|_2.
\end{equation}
Thus, when $T$ is finite we can see that
\begin{equation}
d^{(\infty)}_n (T,{\mathcal S}) = \inf_{\mathcal{A} \in \Gamma_n (\mathcal{S})} ~ \left \| {\bf e}_{\mathcal{A}} \right \|_{\infty},
\end{equation}
and the least squares approximation error over all subspaces in $\Gamma_n (\mathcal{S})$ is given by
\begin{equation}
d^{(2)}_n (T,{\mathcal S}) = \inf_{\mathcal{A} \in \Gamma_n (\mathcal{S})} ~ \left \| {\bf e}_{\mathcal{A}} \right \|_2.
\end{equation}
These two quantities can be seen as extreme instances of the infinite family of approximation errors given by
\begin{equation}
d^{(p)}_n (T,{\mathcal S}) := \inf_{\mathcal{A} \in \Gamma_n (\mathcal{S})} ~ \left \| {\bf e}_{\mathcal{A}} \right \|_p,
\end{equation}
for any parameter $2\leq p\leq \infty$. 
Note that, analogously to \eqref{def:EKnW}, one has
\begin{equation}
d^{(p)}_n (T,{\mathcal S}) := \inf_{\mathcal{A} \in \Gamma_n (\mathcal{S})} ~ d^{(p)}(T,\mathcal{A}),
\end{equation}
where 
\begin{equation}
d^{(p)}(T,\mathcal{A}):=\left \| {\bf e}_{\mathcal{A}} \right\|_p.
\end{equation}
Finally, as above, we note that there will always be at least one optimal affine subspace, $\mathcal{A}_{\rm opt} \in  \Gamma_n (\mathcal{S})$, with
\begin{equation}
d^{(p)}(T,\mathcal{A}_{\rm opt}) = d^{(p)}_n (T,{\mathcal S})
\end{equation}
when $T$ is finite.

\subsection{Symmetry, Ellipsoids, and Properties of $n$-widths}

Let $P = \{ {\bf p}_1, \dots, {\bf p}_M \} \subset \mathbbm{R}^N$, and define 
\begin{equation}
\bar{\bf p} :=  \frac{1}{M} \cdot \sum^M_{j=1} {\bf p}_j.
\end{equation}
We will let $\bar{P} \subset \mathbbm{R}^N$ denote the following symmetrized translation of $P$,
\begin{equation}
\bar{P} := (P - \bar{\bf p} ) \cup (\bar{\bf p} - P) \cup \{ {\bf 0}  \}:= \left \{ {\bf p}_j - \bar{\bf p}  ~\big |~ {\bf p}_j \in P \right \} \cup \left \{ \bar{\bf p}  - {\bf p}_j ~\big |~ {\bf p}_j \in P \right \} \cup \{ {\bf 0}  \}.
\end{equation}
We will say that $P$ is \textit{symmetric} if and only if $P = \bar{P}$.  Furthermore, we will denote the convex hull of $P$ by ${\rm CH}(P)$.  The following theorem due to Fritz John \cite{john1948extremum} guarantees the existence of an ellipsoid that approximates ${\rm CH} \left( \bar{P} \right)$ well.
\begin{thm}[John]
Let $K \subset \mathbbm{R}^N$ be a compact and convex set with nonempty interior that is symmetric about the origin (so that $K = -K$).  Then, there is an ellipsoid centered at the origin, $\mathcal{E} \subset \mathbbm{R}^N$, such that $\mathcal{E} \subseteq K \subseteq \sqrt{N} \cdot \mathcal{E}$.
\label{thm:John}
\end{thm}

Given $P \subset \mathbbm{R}^N$, an ellipsoid which is nearly as good an approximation to ${\rm CH} \left( \bar{P} \right)$ as the ellipsoid guaranteed by Thoerem~\ref{thm:John} can be computed in polynomial time (see, e.g., \cite{khachiyan1996rounding,kumar2005minimum,Todd20071731}).  More specifically, one can compute an ellipsoid $\mathcal{E}$ such that $\mathcal{E} \subseteq {\rm CH} \left( \bar{P} \right) \subseteq \sqrt{(1 + \epsilon)N} \cdot \mathcal{E}$ in $O(MN^2(\log N + 1/\epsilon) )$-time for any $\epsilon \in (0, \infty)$ \cite{Todd20071731}.  Finally, in the following Lemma, we summarize a few facts concerning the $n$-widths of finite sets, convex hulls, and ellipsoids that will be useful for establishing our results (proofs are included in Appendix~\ref{lem1Appendix} for the sake of completeness).

\begin{lem}
Let $P = \{ {\bf p}_1, \dots, {\bf p}_M \} \subset \mathbbm{R}^N$, and $\mathcal{E} \subset \mathbbm{R}^N$ be the ellipsoid 
$$\left\{ {\bf x} \in \mathbbm{R}^N ~\big|~ {\bf x}^T Q {\bf x} \leq 1 \right\},$$
where $Q \in \mathbbm{R}^{N \times N}$ is symmetric and positive definite.  Then,
\begin{enumerate}
\item $d^{(\infty)}_n \left(P - {\bf x }, \mathbbm{R}^N \right) = d^{(\infty)}_n \left(P, \mathbbm{R}^N \right)$ for all ${\bf x} \in \mathbbm{R}^N$, and $n = 1, \dots, N$.
\item $\bar{P}$ will have an optimal $n$-dimensional subspace (i.e., with ${\bf a}_{\mathcal{A}_{\rm opt}} = {\bf 0}$) for all $n = 1, \dots, N$.
\item $d^{(\infty)}_n \left(\bar{P}, \mathbbm{R}^N \right) \leq 2 \cdot d^{(\infty)}_n \left(P, \mathbbm{R}^N \right)$ for all $n = 1, \dots, N$.
\item $d^{(\infty)}_n (B, \mathbbm{R}^N) \leq d^{(\infty)}_n (C, \mathbbm{R}^N)$ for all $B \subseteq C \subset \mathbbm{R}^N$, and $n = 1, \dots, N$.
\item $d^{(\infty)}_n ({\rm CH}(P), \mathbbm{R}^N) = d^{(\infty)}_n (P, \mathbbm{R}^N)$ for all $n = 1, \dots, N$.
\item $d^{(\infty)}_n (\mathcal{E}, \mathbbm{R}^N) = \sqrt{\frac{1}{\sigma_{N-n+1} (Q)}}$ for all $n = 1, \dots, N$.  Consequently, an optimal $n$-dimensional subspace for $\mathcal{E}$ is spanned by the eigenvectors of $Q$ associated with $\sigma_N (Q), \dots, \sigma_{N-n+1}(Q)$.
\end{enumerate}
\label{lem:widthBasics}
\end{lem}

We will assume hereafter, without loss of generality, that $P = \{ {\bf p}_0, \dots, {\bf p}_M \} \subset \mathbbm{R}^N$ both spans $\mathbbm{R}^N$ and is symmetric.  If $P$ initially does not span $\mathbbm{R}^N$, we will replace each element of $P$ with the coordinates of its orthogonal projection into the span of $P$, reducing $N$ accordingly.  Any such change of basis for $P$ will lead to no loss of accuracy in our solution.  If $P$ is not symmetric we will approximate $\bar{P}$ by a subspace instead, noting that a translation of our approximating subspace for $\bar{P}$ will still approximate $P$ well by parts $(1) - (4)$ of Lemma~\ref{lem:widthBasics}.  Finally, we will assume hereafter that ${\bf p}_0 = {\bf 0}$.

\section{Dimensionality Reduction Results}
\label{sec:dimReduct}

In this section we establish our main theorems regarding dimensionality reduction.  As we shall see, the main idea behind the proofs of both Theorems~\ref{thm:EmbedWidth} and~\ref{thm:EmbedWidthlp} below is to use fast existing least-squares methods in order to quickly approximate the point set $P$ in a greedy fashion.  To see how this works, note that $P$'s best-fit least squares subspace will generally fail to approximate all of $P$ to within $d_n^{(p)} \left(P, \mathbbm{R}^N \right)$-accuracy when $p > 2$.  However, it \textit{will} generally approximate a large fraction of $P$ sufficiently well.  Furthermore, we can easily tell which portion of $P$ is approximated best.  

Hence, we may employ a divide-and-concur approach:  we $(i)$ approximate $P$ with its best-fit least squares subspace, $(ii)$ identify the half of its points fit the best, $(iii)$ remove them from $P$, and then $(iv)$ repeat the process again on the remaining portion of $P$.  After $O(\log M)$ repetitions we end up with a collection of at most $O(\log M)$ least squares subspaces whose collective span is guaranteed to contain a near-optimal $n$-dimensional approximation to all of $P$ with respect to $d_n^{(p)} \left(P, \mathbbm{R}^N \right)$.  We are now ready to begin proving our results.

\begin{lem}\label{lem:orderstat}
Let $P = \{{\bf p}_0:= {\bf 0}, {\bf p}_1, \dots, {\bf p}_{M} \} \subset \mathbbm{R}^N$ be symmetric, $n \in \{ 1, \dots, N \}$, and $p \in (2,\infty]$. Then there is an $O\left( M N^2 \right)$-time\footnote{We assume here that $M \geq N \geq \log M$.  We also note that this runtime complexity can be improved substantially by utilizing randomized low-rank approximation algorithms.  See Remark~\ref{rem:randSVD} below.} algorithm which outputs 
an $n$-dimensional subspace $\mathcal{S} \subset \mathbbm{R}^N$ such that for $m\in\{1,\dots,M\}$ one has 
\begin{equation}
\| {\bf p}_{l_m} - \Pi_{\mathcal S} {\bf p}_{l_m} \|^2_2 \leq \frac{M^{1-\frac{2}{p}}}{M-m+1}  \cdot \big(d_n^{(p)} \left(P, \mathbbm{R}^N \right)\big)^2, 
\label{eqn:lembound}
\end{equation}
where the $\ell_i>0$, $i=1,\dots, M$, are chosen to satisfy 
\begin{equation}
0 = \| {\bf p}_{0} - \Pi_{\mathcal S} {\bf p}_{0} \|_2 \leq \| {\bf p}_{l_1} - \Pi_{\mathcal S} {\bf p}_{l_1} \|_2 \leq  \| {\bf p}_{l_2} - \Pi_{\mathcal S} {\bf p}_{l_2} \|_2 \leq \dots \leq  \| {\bf p}_{l_{M}} - \Pi_{\mathcal S} {\bf p}_{l_{M}} \|_2.
\label{eqn:POrdering}
\end{equation}
\end{lem}

\noindent \textit{Proof:}  Denote the matrix whose columns are the points in $P$ by $X \in \mathbbm{R}^{N \times M}$.  That is, let
\begin{equation}
X := \left( {\bf p}_1, \dots, {\bf p}_{M} \right).
\label{DefX}
\end{equation}
Let $\mathcal{A}^{(p)}_{\rm opt} \in \Gamma_n(\mathbbm{R}^D)$ be an optimal $n$-dimensional subspace for $P$ satisfying 
\begin{equation}
d (P,\mathcal{A}^{(p)}_{\rm opt}) = d^{(p)}_n \left( P, \mathbbm{R}^N \right).
\end{equation}
It is not difficult to see that we will have $X = Y + E$, where $Y,E \in \mathbbm{R}^{N \times M}$ have the following properties:  the column span of $Y$ is contained in $\mathcal{A}^{(p)}_{\rm opt}$, 
and the vector $\bf{e}$  whose entries are the $\ell^2$-norms of the columns of $E$  has $\ell^p$-norm at most $d^{(p)}_n \left( P, \mathbbm{R}^N \right)$.  It follows from H\"older's inequality that
\begin{equation}
\sum^{\min(N,M)}_{l = 1} \sigma^2_l (E) =  \| E \|^2_F =\|{\bf{e}}\|^2_2 \leq \|{\bf{e}}\|^2_p \|{\mathbb{I}}\|_{1+\frac{2}{p-2}}= M^{1-\frac{2}{p}} \cdot \big(d_n^{(p)} \left(P, \mathbbm{R}^N \right)\big)^2,
\label{eq:EProp}
\end{equation}
where $\mathbb{I} \in \mathbbm{R}^M$ is the vector whose entries are all one.
Note that $Y$ has rank at most $n$ so that 
\begin{equation}
\sigma_{n+1}(Y) = \dots = \sigma_{\min(N,M)}(Y) = 0.
\end{equation}
Applying Theorem~\ref{Thm:WBounds} we now learn that 
\begin{equation}
\sigma_{n+l}(X) \leq \sigma_{l}(E)
\label{eq:MboundedbyE}
\end{equation}
for all $l \in \{1, \dots, N-n \}$.  

Let $X_n$ be the best rank $n$ approximation to $X$ with respect to Frobenius norm,
\begin{equation}
X_n := \argmin_{\substack{ L \in \mathbbm{R}^{N \times M}\\{\rm rank}~L=~n}} \| X - L \|_{F}.
\label{DefXn}
\end{equation}
Let $\mathcal{S}$ be the $n$-dimensional subspace spanned by the columns of $X_n$.  We have that
\begin{equation}
\| X - X_n \|^2_F = \sum^{\min(N,M)}_{l=n+1} \sigma^2_l(X) \leq M^{1-\frac{2}{p}} \cdot \big(d_n^{(p)} \left(P, \mathbbm{R}^N \right)\big)^2
\label{equ:GrossError}
\end{equation}
due to \eqref{eq:EProp} and \eqref{eq:MboundedbyE}.  Thus, for each positive integer $k$ there can be at most $k$ 
(nonzero) columns of $X$, ${\bf p}_j \in P$, with the property that
\begin{equation}
\| {\bf p}_{j} - \Pi_{\mathcal S} {\bf p}_{j} \|^2_2 \geq \frac{M^{1-\frac{2}{p}}}{k}  \cdot \big(d_n^{(p)} \left(P, \mathbbm{R}^N \right)\big)^2.
\label{equ:CombError}
\end{equation}
Setting $k = M-m+1$, we see that \eqref{eqn:lembound} must hold in order for 
\begin{equation}
\sum^M_{j = m} \| {\bf p}_{l_j} - \Pi_{\mathcal S} {\bf p}_{l_j} \|^2_2 \leq \sum^M_{j = 1} \| {\bf p}_{l_j} - \Pi_{\mathcal S} {\bf p}_{l_j} \|^2_2 \leq \| X - X_n \|^2_F \leq M^{1-\frac{2}{p}} \cdot \big(d_n^{(p)} \left(P, \mathbbm{R}^N \right)\big)^2
\end{equation}
to hold (i.e., in order for \eqref{equ:GrossError} to hold) .

To finish, we note that the subspace $\mathcal{S}$ above is spanned by the $n$ left singular vectors of $X$ associated with its $n$ largest singular values. These can be computed deterministically in $O \left( N M \cdot  \min \{ N, M \}  \right)$-time as part of the full singular value decomposition of $X$, although significantly faster (randomized) approximation algorithms exist (see, e.g., \cite{watkins2007matrix,halko2011finding}). The stated runtime complexity follows given our assumption that $M \geq N \geq \log M$.
\qed

\begin{lem}
Let $\xi \in (1,\infty)$, $P = \{ {\bf p}_0:= {\bf 0}, {\bf p}_1, \dots, {\bf p}_M \} \subset \mathbbm{R}^N$ be symmetric, and $n \in \{ 1, \dots, N \}$.  Then, there is an $O\left( M N^2 \right)$-time\footnote{Again, we assume that $M \geq N \geq \log M$.} algorithm which outputs both an $n$-dimensional subspace $\mathcal{S} \subset \mathbbm{R}^N$, and a symmetric subset $P' \subset P$ with $| P' | \geq \lceil (1 - 1/ \xi)M \rceil + 1$, such that
\begin{equation}
d^{(\infty)} (P',{\mathcal S}) < \sqrt{\xi} \cdot d^{(\infty)}_n \left( P, \mathbbm{R}^N \right).  
\label{lem1goal}
\end{equation}
\label{lem:Froblem}
\end{lem}
\noindent \textit{Proof:} 
We first order the nonzero elements of $P$ according to \eqref{eqn:POrdering}, and then set
\begin{equation}
P' := \left\{ {\bf p}_{0}, {\bf p}_{l_1}, {\bf p}_{l_2}, \dots, {\bf p}_{l_{\lceil (1 - 1/\xi) M \rceil}} \right\} \subset P.
\label{Ppdef}
\end{equation}
If $P'$ is not symmetric, continue to add additional points from $P$ until it is (i.e., by adding the negation of each current point in $P'$ to $P'$).  Applying Lemma~\ref{lem:orderstat} with $m=\lceil (1 - 1/\xi) M \rceil$, we see that
\begin{equation}
\| {\bf p}_{\lceil (1 - 1/\xi) M \rceil} - \Pi_{\mathcal S} {\bf p}_{\lceil (1 - 1/\xi) M \rceil} \|^2_2 \leq \xi \cdot \left(d^{(\infty)}_n \left(P, \mathbbm{R}^N \right)\right)^2.
\end{equation}
Thus there can be at most $\lfloor M / \xi \rfloor$ (nonzero) columns of $X$, ${\bf p}_j \in P$, with the property that
\begin{equation}
\| {\bf p}_{j} - \Pi_{\mathcal S} {\bf p}_{j} \|^2_2 \geq \xi \cdot \left(d^{(\infty)}_n \left(P, \mathbbm{R}^N \right)\right)^2.
\label{equ:CombError}
\end{equation}
By the ordering \eqref{eqn:POrdering}, the associated indices $j$ must be contained in $\{\ell_{\lceil (1 - 1/\xi) M \rceil + 1}, \dots, \ell_{M}\}$, hence  $P' \subset P$ will satisfy \eqref{lem1goal}.  

By Lemma~\ref{lem:orderstat}, a suitable set $\mathcal{S}$ can be found in $O \left( N M \cdot  \min \{ N, M \}  \right)$-time.  Having computed (the singular value decomposition of) $X_n$, the ordering in \eqref{eqn:POrdering} can then be determined in $O(N M + M \log M)$-time.  Finally, the symmetry of $P'$ can be ensured in $O(N M \log M)$-time by, e.g., ordering the points of $P'$ lexicographically, and then performing a binary search for the negation of each point in order to ensure its inclusion. The stated runtime complexity follows given our assumption that $M \geq N \geq \log M$. \qed

\begin{rem}
The runtime complexity quoted in Lemma~\ref{lem:orderstat} and consequently also Lemma~\ref{lem:Froblem} and Lemma~\ref{lem:Froblemlp} is dominated by the time required to compute $X_n$ \eqref{DefXn} via the full singular value decomposition of $X$ \eqref{DefX}.  However, computing $X_n$ this way is computationally wasteful when $n \ll \min \{N, M \}$.  Note that it suffices to find a $O(n)$-dimensional matrix, $\tilde{X}_n \in \mathbbm{R}^{N \times M}$, with the property that
\begin{equation}
\| X - \tilde{X}_n \|_F \leq C \cdot \| X - X_n \|_F
\label{ApproxLowRank}
\end{equation}
for a suitably small constant $C$.  Taking $\tilde{\mathcal{S}}$ to be the column span of $\tilde{X}_n$ in the proof of Lemma~\ref{lem:Froblem} then produces a similarly sized subset $P' \subset P$ satisfying $d^{(\infty)} (P',\tilde{\mathcal S}) \leq C\sqrt{\xi} \cdot d^{(\infty)}_n \left( P, \mathbbm{R}^N \right).$  A tremendous number of methods have been developed for rapidly computing an $\tilde{X}_n$ as above (see, e.g., \cite{watkins2007matrix,halko2011finding}).  In particular, we note here that there exists a modest absolute constant $C \in \mathbbm{R}^+$ such that a randomly constructed matrix $\tilde{X}_n$ of rank $\max \{2n, 7\}$ will satisfy \eqref{ApproxLowRank} with probability $> 0.9$.\footnote{See Theorem 10.7 from \cite{halko2011finding} for more details concerning the constant $C$, etc..  Also, note that the probability of satisfying \eqref{ApproxLowRank} can be boosted as close to $1$ as desired by constructing several different $\tilde{X}_n$ matrices independently, and then choosing the 
most accurate one.}  
Furthermore, this matrix can always be constructed in $O(N M n + N n^2)$-time.\\
\label{rem:randSVD}
\end{rem}

\begin{lem}
Let $p \in (2,\infty)$, $\xi \in \left(1,M/2 \right]$, $P = \{ {\bf p}_1, \dots, {\bf p}_M \} \subset \mathbbm{R}^N$ be symmetric, and $n \in \{ 1, \dots, N \}$.  Then, there is an $O\left( M N^2 \right)$-time\footnote{Again, we assume that $M \geq N \geq \log M$.} algorithm which outputs both an $n$-dimensional subspace $\mathcal{S} \subset \mathbbm{R}^N$, and a symmetric subset $P' \subset P$ with $| P' | \geq \left\lceil \left(1 - 1/\xi \right) M \right\rceil + 1$, such that 
\begin{equation}
d^{(p)} (P',{\mathcal S}) 
\leq C \sqrt{\xi} \cdot d^{(p)}_n \left( P, \mathbbm{R}^N \right).
\label{lemgoallp}
\end{equation}
\label{lem:Froblemlp}
\end{lem}

\noindent \textit{Proof:}  We again order the nonzero elements of $P$ according to \eqref{eqn:POrdering}, and then define $P'$ as above in \eqref{Ppdef}.
From Lemma~\ref{lem:orderstat} with $m = \left\lceil \left(1 - 1/\xi \right) M \right\rceil$ we obtain that 
\begin{align}
(d^{(p)} (P',{\mathcal S}))^p&= \sum_{j=1}^m \| {\bf p}_{\ell_j} - \Pi_{\mathcal S} {\bf p}_{\ell_j} \|^p_2 \\
& \leq \sum_{j=1}^{m}
\left(\frac{M^{1-\frac{2}{p}}}{M-j+1}  \cdot \left(d_n^{(p)} \left(P, \mathbbm{R}^N \right)\right)^2\right)^{p/2}\\
& = M^{\frac{p}{2}-1} \left(d_n^{(p)} \left(P, \mathbbm{R}^N \right)\right)^p \sum^M_{j = M-m+1} j^{-p/2}\\
&\leq M^{\frac{p}{2}-1} \left(d^{(p)}_n \left( P, \mathbbm{R}^N \right)\right)^p \int_{M-m}^M x^{-p/2} dx \label{eq:intest}\\
&= \frac{\left( 1- \frac{m}{M} \right)^{1-\frac{p}{2}} -1}{\frac{p}{2}-1} \left(d^{(p)}_n \left( P, \mathbbm{R}^N \right)\right)^p. \label{EndCalcs}
\end{align}

Set $\delta := m/M - \left(1 - 1/ \xi \right) < 1/M$.  It is not difficult to see that $1/\xi - \delta \in (0,1)$ since $\xi \in \left(1,M/2 \right]$.  Thus, 
\begin{equation}
\left( \left(1/\xi \right) - \delta \right)^{1-\frac{p}{2}} \leq  \left( \frac{\xi}{1-\xi/M} \right)^{\frac{p}{2}-1} \leq (2 \xi)^{\frac{p}{2}-1}, 
\end{equation}
which now allows us to bound \eqref{EndCalcs} as follows:
\begin{equation}
(d^{(p)} (P',{\mathcal S}))^p \leq \frac{\left( 1- \frac{m}{M} \right)^{1-\frac{p}{2}} -1}{\frac{p}{2}-1} \left(d^{(p)}_n \left( P, \mathbbm{R}^N \right)\right)^p < \frac{(2 \xi)^{\frac{p}{2}-1} - 1}{\frac{p}{2}-1} \cdot\left(d^{(p)}_n \left( P, \mathbbm{R}^N \right)\right)^p.
\end{equation}
This last expression yields the first inequality in \eqref{lemgoallp}, as desired.  For large $p$, the lemma directly follows from the asymptotics, for $p\approx 2$ from l'Hospital's rule.  As the set $P'$ is constructed in the same way as in the proof of Lemma~\ref{lem:Froblem}, the runtime analysis given there carries over directly. \qed

\begin{rem}
Note that the ordered distances \eqref{eqn:POrdering} between the points in $P$ and the subspace $\mathcal{S}$ from Lemma~\ref{lem:Froblem} satisfy
\begin{equation}
\left \| {\bf p}_{l_{\lceil (1 - 1/\xi) M \rceil}} - \Pi_{\mathcal S} {\bf p}_{l_{\lceil (1 - 1/\xi) M \rceil}} \right \|_2 \leq \sqrt{\xi} \cdot d_n \left( P, \mathbbm{R}^N \right).
\end{equation}
We can use this information to bound $d_n \left( P, \mathbbm{R}^N \right)$ from above and below.  Set
\begin{equation}
\alpha := \frac{\| {\bf p}_{l_{M}} - \Pi_{\mathcal S} {\bf p}_{l_{M}} \|_2}{\left \| {\bf p}_{l_{\lceil (1 - 1/\xi) M \rceil}} - \Pi_{\mathcal S} {\bf p}_{l_{\lceil (1 - 1/\xi) M \rceil}} \right \|_2}.
\label{Defalpha}
\end{equation}
We now have
\begin{equation}
 d_n \left( P, \mathbbm{R}^N \right) \leq \| {\bf p}_{l_{M-1}} - \Pi_{\mathcal S} {\bf p}_{l_{M-1}} \|_2  = \alpha \cdot \left \| {\bf p}_{l_{\lceil (1 - 1/\xi) M \rceil}} - \Pi_{\mathcal S} {\bf p}_{l_{\lceil (1 - 1/\xi) M \rceil}} \right \|_2 \leq \alpha \sqrt{\xi} \cdot d_n \left( P, \mathbbm{R}^N \right).
 \end{equation}
Thus, computing $\alpha$ allows us to estimate  $d_n \left( P, \mathbbm{R}^N \right)$.  If $\alpha$ is sufficiently small, $\mathcal{S}$ will itself be a passible approximation to an optimal subspace $\mathcal{A}_{\rm opt}$.  Similarly, if $P' \subset P$ and $\mathcal{S}$ from Lemma~\ref{lem:Froblemlp} satisfy 
\begin{equation}
d^{(p)}(P,\mathcal{S})  \leq \alpha \cdot d^{(p)}(P',\mathcal{S})
\end{equation}
for a modest $\alpha \in \mathbbm{R}^+$, then we may infer that $\mathcal{S}$ is a near-optimal subspace for $P$.
\label{rem:EstKolWidth}
\end{rem}

Lemmas~\ref{lem:Froblem} and~\ref{lem:Froblemlp} now allow us to establish the main results of this section.

\begin{thm}
Let $\xi \in (1,\infty)$, $P = \{ {\bf p}_1, \dots, {\bf p}_M \} \subset \mathbbm{R}^N$ be symmetric, and $n \in \{ 1, \dots, N \}$.  Then, there is an $O \left( \frac{\xi}{\xi - 1} \cdot M N^2  + N \cdot n^2 \log^2_\xi M \right)$-time algorithm which outputs an at most $(n \cdot \lceil \log_{\xi} M \rceil)$-dimensional subspace $\mathcal{S} \subset \mathbbm{R}^N$ with
\begin{equation}
d^{(\infty)}_n (P,{\mathcal S}) \leq \left(1+\sqrt{\xi} \right) \cdot d^{(\infty)}_n \left( P, \mathbbm{R}^N \right).
\label{thm3goal}
\end{equation}
\label{thm:EmbedWidth}
\end{thm}
\noindent \textit{Proof:} 
Let $\mathcal{S} \subset \mathbbm{R}^D$ be an $\tilde{n}$-dimensional subspace with $\tilde{n} \geq n$, and $\mathcal{A} \in  \Gamma_n(\mathbbm{R}^D)$.  We have that
\begin{equation}
d^{(\infty)}_n(P,\mathcal{S}) \leq \max_{{\bf p}_j \in P} \| {\bf p}_j -  \Pi_{\mathcal S} \Pi_{\mathcal A} {\bf p}_j \|_2 \leq \max_{{\bf p}_j \in P}  \| {\bf p}_j -  \Pi_{\mathcal S} {\bf p}_j \|_2 + \max_{{\bf p}_j \in P}  \| {\bf p}_j -  \Pi_{\mathcal A} {\bf p}_j \|_2. 
\end{equation}
The fact that this holds for all $\mathcal{A} \in  \Gamma_n(\mathbbm{R}^D)$ now immediately implies that 
\begin{equation}
d^{(\infty)}_n(P,\mathcal{S}) \leq d^{(\infty)}(P,\mathcal{S}) + d^{(\infty)}_n \left( P, \mathbbm{R}^N \right).
\label{eqn:GoodVNeeded}
\end{equation}
It remains to make a good choice for the subspace $\mathcal{S}$.  More precisely, we would like to find a subspace $\mathcal{S}$ with $d^{(\infty)}(P,\mathcal{S}) \leq \sqrt{\xi} \cdot d^{(\infty)}_n \left( P, \mathbbm{R}^N \right)$ so that we can obtain \eqref{thm3goal} from \eqref{eqn:GoodVNeeded}.

Appealing to Lemma~\ref{lem:Froblem}, we note that we can find a sufficiently accurate $n$-dimensional subspace, $\mathcal{S}^1$, for a large symmetric subset $P' \subset P$ with $|P'| \geq \lceil (1 - 1/ \xi)M \rceil + 1$.  It remains to find a similarly accurate subspace for the rest of $P$.  Set $P_2 := P - P' \cup \{ 0 \}$, noting that $P_2$ will be a symmetric point set with $|P_2| \leq M / \xi $.  We may now apply Lemma~\ref{lem:Froblem} to $P_2$ in order to find a second $n$-dimensional subspace, $\mathcal{S}^2$, which approximates all but at most $M / \xi^2$ elements of $P_2$ to within the desired $\sqrt{\xi} \cdot d^{(\infty)}_n \left( P, \mathbbm{R}^N \right)$-accuracy.  More generally, we can see that iterating Lemma~\ref{lem:Froblem} at most $\lceil \log_\xi M \rceil$-times in this fashion will produce a collection of at most $\lceil \log_\xi M \rceil$ different $n$-dimensional subspaces, $\mathcal{S}^1, \dots, \mathcal{S}^{\lceil \log_\xi M \rceil}$, which will collectively approximate all of 
$P$ to the 
desired $\sqrt{\xi} \cdot d^{(\infty)}_n \left( P, \mathbbm{R}^N \right)$-accuracy.  We now set
\begin{equation}
\mathcal{S} := {\rm span} \left(\mathcal{S}^1\cup \dots \cup \mathcal{S}^{\lceil \log_\xi M \rceil} \right).
\end{equation}

It is not difficult to see that $\mathcal{S}$ will be at most $(n \cdot \lceil \log_{\xi} M \rceil)$-dimensional.  Furthermore, the at most $\lceil \log_\xi M \rceil$ applications of 
Lemma~\ref{lem:Froblem} will induce a runtime of complexity of
\begin{equation}
O\left( \sum^{\lceil \log_\xi M \rceil -1 }_{j=0}  \frac{N M \cdot \min \{ N, M/ \xi^j \} }{\xi^j} \right) = O \left( \frac{\xi}{\xi - 1} \cdot M N^2 \right).
\end{equation}
Finally, we note that an orthonormal basis for $\mathcal{S}$ can be computed in $O \left( N \cdot n^2 \log^2_\xi M \right)$-time via Gram--Schmidt.  The stated result follows.\qed

\begin{thm}
Let $\xi \in (1,\infty)$, $P = \{ {\bf p}_1, \dots, {\bf p}_M \} \subset \mathbbm{R}^N$ be symmetric, and $n \in \{ 1, \dots, N \}$. Then, there is an $O \left( \frac{\xi}{\xi - 1} \cdot M N^2  + N \cdot n^2 \log^2_\xi M \right)$-time algorithm which outputs an at most $(n \cdot \lceil \log_{\xi} M \rceil)$-dimensional subspace $\mathcal{S} \subset \mathbbm{R}^N$ such that one has for an absolute constant $C$, simultaneously for all $2<p<\infty$,
\begin{equation}
d_n^{(p)} (P,{\mathcal S}) \leq \left(1+ C \lceil \log_\xi m\rceil^{1/p} \sqrt{\xi} \right) \cdot d_n^{(p)} \left( P, \mathbbm{R}^N \right). 
\label{thm4goal}
\end{equation}
\label{thm:EmbedWidthlp}
\end{thm}

\noindent \textit{Proof:}  Let $\mathcal{S} \subset \mathbbm{R}^D$ be an $\tilde{n}$-dimensional subspace with $\tilde{n} \geq n$, and $\mathcal{A} \in  \Gamma_n(\mathbbm{R}^D)$.  We have that
\begin{equation*}
d_n^{(p)}(P,\mathcal{S}) \leq \left(\sum_{{\bf p}_j \in P} \| {\bf p}_j -  \Pi_{\mathcal S} \Pi_{\mathcal A} {\bf p}_j \|_2^p\right)^{1/p} \leq \left(\sum_{{\bf p}_j \in P}  \| {\bf p}_j -  \Pi_{\mathcal S} {\bf p}_j \|_2^p\right)^{1/p} + \left(\sum_{{\bf p}_j \in P}  \| {\bf p}_j -  \Pi_{\mathcal A} {\bf p}_j \|^p_2\right)^{1/p}. 
\end{equation*}
The fact that this holds for all $\mathcal{A} \in  \Gamma_n(\mathbbm{R}^D)$ now again implies that 
\begin{equation}
d_n^{(p)}(P,\mathcal{S}) \leq d^{(p)}(P,\mathcal{S}) + d^{(p)}_n \left( P, \mathbbm{R}^N \right).
\label{eqn:GoodVNeeded_p}
\end{equation}
The  subspace $\mathcal{S}$ is chosen in the same way as in the proof of Theorem~\ref{thm:EmbedWidth}. That is, it is given as the union of $\lceil \log_\xi m\rceil$ recursively constructed subsets $\mathcal{S}^1,\dots,\mathcal{S}^{\lceil \log_\xi m\rceil}$. As both Lemma~\ref{lem:Froblem} and Lemma~\ref{lem:Froblemlp}, the former of which motivates the construction of $\mathcal{S}$, restrict $P$ to the same subset $P'$, we can conclude for the partition $P=\bigcup_{j=1}^{\lceil \log_\xi m\rceil} P_i$ of Theorem~\ref{thm:EmbedWidth}, that each $\mathcal{S}^i$ approximates $P_i$ also in the sense of $d^{(p)}$. That is,
\begin{equation}
d^{(p)} (P_i,{\mathcal S}^i) \leq  C \sqrt{\xi} \cdot d^{(p)}_n \left( P, \mathbbm{R}^N \right).
\end{equation}
Combining the contributions of the different $P_i$, we obtain using Lemma~\ref{lem:Froblemlp}
\begin{equation}
(d^{(p)} (P,{\mathcal S}))^p \leq  \sum_{j=1}^{\lceil \log_\xi m\rceil}(d^{(p)} (P_i,{\mathcal S}))^p \leq \sum_{j=1}^{\lceil \log_\xi m\rceil}(d^{(p)} (P_i,{\mathcal S^i}))^p \leq C^p  \lceil \log_\xi m\rceil \xi^{p/2} \cdot \left( d^{(p)}_n \left( P, \mathbbm{R}^N \right) \right)^p.
\end{equation}

As the construction is the same, the runtime estimate of Theorem~\ref{thm:EmbedWidth} carries over.  The stated result follows.\qed

\begin{rem}
Recalling Remark~\ref{rem:randSVD}, we note that the runtime complexities quoted in both Theorems~\ref{thm:EmbedWidth} and~\ref{thm:EmbedWidthlp} can be reduced by using faster randomized row-rank approximation methods in Lemmas~\ref{lem:Froblem} and~\ref{lem:Froblemlp}, respectively.   Furthermore, we point out that one can use the ideas from Remark~\ref{rem:EstKolWidth} in order to guarantee a, e.g., $2\sqrt{\xi} \cdot d^{(\infty)}_n \left( P, \mathbbm{R}^N \right)$-accurate approximation to $P$ with potentially fewer than $\lceil \log_\xi M \rceil$ applications of Lemma~\ref{lem:Froblem}.  This can be achieved by terminating the iterative applications of Lemma~\ref{lem:Froblem} described in the proof of Theorem~\ref{thm:EmbedWidth} once $\alpha$ from $\eqref{Defalpha}$ falls below $2$.  Similarly, the iterative applications of Lemma~\ref{lem:Froblemlp} described in the proof of Theorem~\ref{thm:EmbedWidthlp} can be terminated without seriously degrading accuracy as soon as $\alpha := d^{(p)}(P,\mathcal{S})
 / d^{(p)}(
P',\mathcal{S})$ falls below a user prescribed threshold.  Finally, it worth noting that the accuracy of Theorem~\ref{thm:EmbedWidth} (and Theorem~\ref{thm:EmbedWidthlp}) can be improved in practice by replacing $P \setminus P'$ with $\left(I - \Pi_{\mathcal{S}}\right) (P \setminus P')$ after each iteration of Lemma~\ref{lem:Froblem} (or Lemma~\ref{lem:Froblemlp}).  This allows subsequent iterations to strictly improve on the progress made in previous iterations.
\end{rem}

\section{A Fast Algorithm for $p = \infty$ Subspace Approximation}
\label{sec:CaseInfRes}

In this section we demonstrate that the dimensionality reduction results developed above can be combined with computational techniques for computing the John ellipsoid of a point set in order to produce a fast approximation algorithm for the $p = \infty$ problem.  The following result establishes the speed and accuracy of this approach.

\begin{thm}
Let $P = \{ {\bf p}_1, \dots, {\bf p}_M \} \subset \mathbbm{R}^N$ be symmetric, and $n \in \{ 1, \dots, N \}$.  Then, one can calculate an $\mathcal{A} \in \Gamma_{n} \left( \mathbbm{R}^N \right)$ with 
\begin{equation}
d^{(\infty)} \left( P, \mathcal{A} \right) \leq C \sqrt{n \cdot \log M} \cdot d^{(\infty)}_n \left( P, \mathbbm{R}^N \right)
\label{Coro1goal}
\end{equation}
in $O \left(M N^2  + M n^2 \cdot \log^2 M \cdot  \log (n \log M)  \right)$-time.  Here $C \in \mathbbm{R}^+$ is an absolute constant.
\label{thm:EllipsplusDimReduct}
\end{thm}

\noindent \textit{Proof:}
Choose $\epsilon \in (0, \infty)$, $\xi \in (1,\infty)$, and let $m := n \lceil \log_\xi M \rceil$.  Use Theorem~\ref{thm:EmbedWidth} to find $\mathcal{S} \in \Pi_{\tilde{m}} \left( \mathbbm{R}^N \right)$ with $n \leq \tilde{m} \leq m$, and let $B_\mathcal{S}$ be the associated orthonormal basis of $\mathcal{S}$.  Project $P$ onto $\mathcal{S}$ to obtain $P' := \Pi_\mathcal{S} P \subset \mathbbm{R}^N$.  We will also work with $P'$ expressed in terms of its $B_\mathcal{S}$ coordinates, $P'' \subset \mathbbm{R}^{\tilde{m}}$.  Compute an ellipsoid $\mathcal{E} := \left\{ {\bf x} ~\big|~ {\bf x}^T Q {\bf x} \leq 1 \right\} \subset \mathbbm{R}^{\tilde{m}}$ such that $\mathcal{E} \subseteq {\rm CH} \left( P'' \right) \subseteq \sqrt{(1 + \epsilon)m} \cdot \mathcal{E}$ in $O \left(M m^2 ( \log m + 1/\epsilon) \right)$-time \cite{Todd20071731}.  Finally, let $\mathcal{A'_E} \subset \mathbbm{R}^{\tilde{m}}$ be the subspace spanned by the $n$ eigenvectors of $Q$ associated with $\sigma_{\tilde{m}} (Q), \dots, 
\sigma_{\tilde{m}-n+1}(Q)$, and $\mathcal{A_E} \subset \mathcal{S} \
\subset \mathbbm{R}^N$ be $\mathcal{A'_E}$ re-expressed as an $n$-dimensional subspace of the span of $B_\mathcal{S}$.  

Choosing $\mathcal{A}'_{\rm opt} \in \Gamma_{n} \left( \mathbbm{R}^{\tilde{m}} \right)$ to satisfy $d^{(\infty)} \left( {\rm CH} \left(P''\right), \mathcal{A}'_{\rm opt} \right) = d^{(\infty)}_n \left( {\rm CH} \left(P''\right) , \mathbbm{R}^{\tilde{m}} \right) = d^{(\infty)}_n \left( P'' , \mathbbm{R}^{\tilde{m}} \right)$, one can see that 
\begin{align}
d^{(\infty)} \left( P', \mathcal{A_E} \right) & = d^{(\infty)} \left( P'',\mathcal{A'_E} \right) \leq d^{(\infty)} \left( {\rm CH} \left(P'' \right),\mathcal{A'_E} \right) \leq d^{(\infty)} \left( \sqrt{(1 + \epsilon)m} \cdot \mathcal{E} ,\mathcal{A'_E} \right)\label{equ:bound1}\\
&=\sqrt{(1 + \epsilon)m} \cdot d^{(\infty)} \left( \mathcal{E} ,\mathcal{A'_E} \right) \leq \sqrt{(1 + \epsilon)m} \cdot d^{(\infty)} \left( \mathcal{E},\mathcal{A}'_{\rm opt} \right)\\
&= \sqrt{(1 + \epsilon)m} \cdot d^{(\infty)} \left( {\rm CH} \left(P'' \right) ,\mathcal{A}'_{\rm opt} \right) = \sqrt{(1 + \epsilon)m} \cdot d^{(\infty)}_n \left( P'' , \mathbbm{R}^{\tilde{m}} \right).
\label{equ:bound2}
\end{align}
where  the inequality in \eqref{equ:bound1} follows from parts $(5)$ and $(6)$ of Lemma~\ref{lem:widthBasics}.
Finally, after noting that $d^{(\infty)}_n \left( P'' , \mathbbm{R}^{\tilde{m}} \right) = d^{(\infty)}_n \left( P' , \mathcal{S} \right)$, we can see that \eqref{equ:bound1}-\eqref{equ:bound2} imply that
\begin{equation}
d^{(\infty)} \left( P', \mathcal{A_E} \right) \leq  \sqrt{(1 + \epsilon)m} \cdot d^{(\infty)}_n \left( P' , \mathcal{S} \right).
\label{eqn:ErrorOnS}
\end{equation}

Choose any $\mathcal{A} \in \Gamma_{n} \left( \mathcal{S} \right)$, thus ensuring that $\Pi_{\mathcal{A}} \Pi_{\mathcal{S}}=\Pi_{\mathcal{A}}$, and then let ${\bf y} \in P$ be such that
\begin{equation}
\left\| \Pi_{\mathcal{S}} {\bf y} - \Pi_{\mathcal{A}} {\bf y} \right\|_2 = \left\| \Pi_{\mathcal{S}} {\bf y} - \Pi_{\mathcal{A}} \Pi_{\mathcal{S}} {\bf y} \right\|_2 = d^{(\infty)} \left( P' , \mathcal{A} \right)\geq d^{(\infty)}_n \left( P' , \mathcal{S} \right). \label{eqn:Abound}
\end{equation}
Choose any ${\bf x} \in P$.  Combining \eqref{eqn:ErrorOnS} and \eqref{eqn:Abound}, we can see that
\begin{equation}
\left\| \Pi_{\mathcal{S}} {\bf x} - \Pi_{\mathcal{A_E}} \Pi_{\mathcal{S}} {\bf x} \right\|^2_2 =  \left\| \Pi_{\mathcal{S}} {\bf x} - \Pi_{\mathcal{A_E}} {\bf x} \right\|^2_2 \leq (1 + \epsilon)m \cdot \left\| \Pi_{\mathcal{S}} {\bf y} - \Pi_{\mathcal{A}} {\bf y}\right\|_2^2
\end{equation} 
which implies that 
\begin{equation}
\left\| \Pi_{\mathcal{S}} {\bf x} - \Pi_{\mathcal{A_E}} {\bf x} \right\|^2_2 + \left\| \Pi_{\mathcal{S}^{\perp}} {\bf x} \right \|_2^2 \leq (1 + \epsilon)m \cdot \left( \left\| \Pi_{} {\bf y} - \Pi_{\mathcal{A}} {\bf y} \right\|_2^2 + \left\| \Pi_{\mathcal{S}^{\perp}} {\bf y} \right \|_2^2 \right) + \left(d^{(\infty)} \left(P,\mathcal{S} \right) \right)^2.
\end{equation} 
Here we used that $ \left\| \Pi_{\mathcal{S}^{\perp}} {\bf x} \right \|_2 = \|{\bf x} -\Pi_{\mathcal{S}} {\bf x}\|_2 \leq  d^{(\infty)} \left(P,\mathcal{S} \right)$.
Thus, again for arbitrary $\mathcal{A}\in \Gamma_{n} \left( \mathcal{S} \right)$,
\begin{align}
\left\| {\bf x} - \Pi_{\mathcal{A_E}} {\bf x} \right\|_2 &\leq \sqrt{(1 + \epsilon)m \cdot \left\| {\bf y} - \Pi_{\mathcal{A}} {\bf y} \right\|_2^2 + \left(d^{(\infty)} \left(P,\mathcal{S} \right) \right)^2}\\ &\leq \sqrt{(1 + \epsilon)m \cdot \left( d^{(\infty)} \left( P, \mathcal{A} \right) \right)^2 + \left(d^{(\infty)} \left(P,\mathcal{S} \right) \right)^2}. \label{eqn:LUB}
\end{align} 
Noting that \eqref{eqn:LUB} holds for all ${\bf x} \in P$ and $\mathcal{A} \in \Gamma_{n} \left( \mathcal{S} \right)$, and recalling that $\mathcal{S}$ was provided by Theorem~\ref{thm:EmbedWidth}, we obtain
\begin{equation}
d^{(\infty)} \left( P, \mathcal{A_E} \right) \leq \sqrt{(1 + \epsilon)m \cdot \big( d^{(\infty)}_n \left( P, \mathcal{S} \right) \big)^2 + \xi \big( d^{(\infty)}_n \left( P, \mathbbm{R}^N \right) \big)^2}.
\end{equation}
Appealing to the statement of Theorem~\ref{thm:EmbedWidth} one last time yields \eqref{Coro1goal}.

The runtime complexity can be accounted for as follows:  Computing $\mathcal{S}$ via Theorem~\ref{thm:EmbedWidth} can be accomplished in $O \left( \frac{\xi}{\xi - 1} \cdot M N^2  + N \cdot n^2 \log^2_\xi M \right)$-time.  Computing $P''$ from $P$ can be done in $O(M N \cdot n \log_{\xi} M)$-time, after which $\mathcal{A'_E}$ can be found in $O \left(M \cdot n^2 \log^2_{\xi} M \cdot \left( \log(n \log_{\xi} M) + 1/\epsilon \right) \right)$-time via \cite{Todd20071731}.  Finally, a basis for $\mathcal{A_E}$ can be computed in $O(N \cdot n^2 \log^2_{\xi} M)$-time once $\mathcal{A'_E}$ is known.  The stated runtime complexity follows.\qed

\begin{rem}
The more precise accuracy bound in terms of the parameters $\epsilon$ and $\xi$ derived in the proof of the theorem predicts that one can find a set $\mathcal{A}$ that satisfies 
 \begin{equation}
 d^{(\infty)} \left( P, \mathcal{A} \right) \leq \Big( \sqrt{(1 + \epsilon)\big(1 + \sqrt{\xi} \big)^2 n  \lceil \log_\xi M \rceil + \xi} \Big) \cdot d^{(\infty)}_n \left( P, \mathbbm{R}^N \right)
 \end{equation}
 in $O \big( \frac{\xi}{\xi - 1} \cdot M N^2  + M n^2 \cdot \log^2_\xi M \cdot \left( \log (n \log_\xi M) + 1 / \epsilon \right) \big)$-time. Choosing $\epsilon$ small and $\xi$ to minimize the accuracy bound to find that one can achieve $C<10$.  Finally, we note that the runtime complexity quoted in Theorem~\ref{thm:EllipsplusDimReduct} can be reduced, along the lines of Remark~\ref{rem:randSVD}, by using a fast randomized least-squares method instead of a deterministic SVD method.
\end{rem}

\section*{Acknowledgements} 

The authors would like to thank Kasturi Varadarajan for the helpful comments and advice he kindly provided to us during the preparation of this manuscript.

\bibliographystyle{abbrv}
\bibliography{Kolm_Width}

\appendix

\section{Proof of Lemma~\ref{lem:widthBasics}}
\label{lem1Appendix}

We present the proof of each part below:
\begin{enumerate}

\item This follows directly from the fact that $d^{(\infty)}(P - {\bf x},\mathcal{A}) = d^{(\infty)}\left(P,\mathcal{A}-\Pi_{\mathcal{S}_\mathcal{A}^\perp}{\bf x} \right)$ for all $\mathcal{A} \in \Gamma_{n} \left( \mathbbm{R}^N \right)$ and ${\bf x} \in \mathbbm{R}^N$.\\

\item Let $\mathcal{A} \in \Gamma_{n} \left( \mathbbm{R}^N \right)$ be such that $d^{(\infty)}(\bar{P},\mathcal{A}) = d^{(\infty)}_n (\bar{P}, \mathbbm{R}^N)$.  Suppose ${\bf a}_{\mathcal{A}}$ is nonzero.  Partition $\bar{P}$ into three parts:

\begin{enumerate}
\item $\bar{P}_1 := \left\{ {\bf p} \in \bar{P}~\big|~ \langle {\bf p}, {\bf a}_{\mathcal{A}} \rangle = 0 \right\}$
\item $\bar{P}_2 := \left\{ {\bf p} \in \bar{P}~\big|~ \langle {\bf p}, {\bf a}_{\mathcal{A}} \rangle > 0 \right\}$
\item $\bar{P}_3 := \left\{ {\bf p} \in \bar{P}~\big|~ \langle {\bf p}, {\bf a}_{\mathcal{A}} \rangle < 0 \right\}$
\end{enumerate}

If ${\bf p} \in P_1$ then $\| {\bf p} - \Pi_{\mathcal{A}} {\bf p} \|_2^2 = \| {\bf p} - \Pi_{\mathcal{S}_\mathcal{A}} {\bf p} \|^2_2 + \| {\bf a}_{\mathcal{A}} \|^2_2$.  This is minimized for all ${\bf p} \in P_1$ when $\| {\bf a}_{\mathcal{A}} \|_2 = 0$.  Next, note that ${\bf p} \in P_3$ if and only if $-{\bf p} \in P_2$, and that ${\bf p} \in P_3$ means $\| {\bf p} - \Pi_{\mathcal{A}} {\bf p} \|_2 > \| (-{\bf p}) - \Pi_{\mathcal{A}} (-{\bf p}) \|_2$.  Thus, we can decrease $d^{(\infty)}(\bar{P},\mathcal{A})$ by making ${\bf a}_{\mathcal{A}}$ shorter (a contradiction).\\

\item Let $\mathcal{A} \in \Gamma_{n} \left( \mathbbm{R}^N \right)$ be such that $d^{(\infty)}(P,\mathcal{A}) = d^{(\infty)}_n (P, \mathbbm{R}^N)$.  We have that 
\begin{align}
\| \bar{\bf p} - {\bf p}_j -\Pi_{\mathcal{S}_\mathcal{A}} \left( \bar{\bf p} - {\bf p}_j \right) \|_2 &= \| {\bf p}_j - \bar{\bf p} -\Pi_{\mathcal{S}_\mathcal{A}} \left( {\bf p}_j - \bar{\bf p} \right) \|_2  = \left \| {\bf p}_j  - \Pi_{\mathcal{S}_\mathcal{A}} {\bf p}_j - \Pi_{\mathcal{S}^\perp_\mathcal{A}} \bar{\bf p} \right \|_2 \\ & \leq \| {\bf p}_j  - \Pi_{\mathcal{A}} {\bf p}_j \|_2 +  \| \bar{\bf p}  - \Pi_{\mathcal{A}} \bar{\bf p} \|_2.
\end{align}
Noting that $\| \bar{\bf p}  - \Pi_{\mathcal{A}} \bar{\bf p} \|_2 \leq d^{(\infty)}(P,\mathcal{A})$ -- see part five below for an analogous calculation -- concludes the proof.\\ 

\item This follows directly from the fact that $d^{(\infty)}(B,\mathcal{A}) \leq d^{(\infty)}(C,\mathcal{A})$ for all $\mathcal{A} \in \Gamma_{n} \left( \mathbbm{R}^N \right)$.\\

\item Part four implies $d^{(\infty)}_n (P, \mathbbm{R}^N) \leq d^{(\infty)}_n ({\rm CH}(P), \mathbbm{R}^N)$ since $P \subseteq {\rm CH}(P)$.  To obtain the other inequality, we recall that every ${\bf x} \in {\rm CH}(P)$ has $\alpha_j \in [0,1]$, $j= 1, \dots, M$, such that
\begin{equation}
{\bf x} = \sum^M_{j=1} \alpha_j \cdot {\bf p}_j,
\end{equation}
and
\begin{equation}
\sum^M_{j=1} \alpha_j = 1.
\end{equation}
Hence, we can see that 
\begin{equation}
\| {\bf x} - \Pi_{\mathcal A} {\bf x} \|_2 = \left\| \sum^M_{j=1} \alpha_j \cdot \left({\bf p}_j - \Pi_{\mathcal{S}_{\mathcal A}} {\bf p}_j - {\bf a}_\mathcal{A} \right)  \right\|_2~\leq~ \sum^M_{j=1} \alpha_j \cdot \| {\bf p}_j - \Pi_{\mathcal A} {\bf p}_j \|_2 ~\leq~ d^{(\infty)}(P,\mathcal{A})
\end{equation}
holds for all ${\bf x} \in {\rm CH}(P)$, and $\mathcal{A} \in \Gamma_{n} \left( \mathbbm{R}^N \right)$.  It now follows that $d^{(\infty)}_n ({\rm CH}(P), \mathbbm{R}^N) \leq d^{(\infty)}_n (P, \mathbbm{R}^N)$.\\

\item Part two tells us that there will be an optimal subspace, since $\mathcal{E}$ is symmetric.  Thus, standard results concerning the $n$-widths of ellipsoids apply  (see, e.g., \cite{kolmogoroff1936beste,kolmogorov2007kolmogorov}).
\end{enumerate}

\end{document}